\documentclass[draftcls,onecolumn,conference,12pt]{IEEEtran}
\usepackage{pifont}
\usepackage{times,amsmath,color, balance,
amssymb,graphicx,epsfig,cite,psfrag,algorithm,multirow,cases,balance,algorithmic}

\usepackage{subfigure}
\usepackage{caption}

\makeatletter
\def\hlinewd#1{%
  \noalign{\ifnum0=`}\fi\hrule \@height #1 \futurelet
   \reserved@a\@xhline}
\makeatother

\DeclareMathOperator*{\argmax}{argmax}
\IEEEoverridecommandlockouts
\begin{document}

\title{Indoor Localization Algorithm For Smartphones}
%\author{\\%
%
%}
%\maketitle \thispagestyle{empty}

\author{Kaiqing Zhang$^{*}$, Hong Hu$^{*}$, Wenhan Dai$^{\dagger}$, Yuan Shen$^\ddagger$ and Moe Z. Win$^{\dagger}$\\
$^{*}$Department of Automation, Tsinghua University, Beijing 100084, China\\
$^{\dagger}$Laboratory for Information and Decision Systems, \\Massachusetts Institute of Technology, Cambridge, MA 02139\\
$^{\ddagger}$Department of Electronic Engineering, Tsinghua University, Beijing 100084, China\\
Email: \{zkq11@mails.tsinghua.edu.cn, hhtsinghua100@126.com, whdai@mit.edu,  \\shenyuan\_ee@tsinghua.edu.cn, moewin@mit.edu\}
\thanks {$^{*}$ The first two authors contribute equally to this work.}
}
\maketitle \thispagestyle{empty}

\begin{abstract}
Increasing sources of sensor measurements and prior knowledge have become available for indoor localization on smartphones. How to effectively utilize these sources for enhancing localization accuracy is an important yet challenging problem. In this paper, we present an area state-aided localization algorithm that exploits various sources of information.
Specifically, we introduce the concept of \emph{area state} to indicate the area where the user is on an indoor map. The position of the user is then estimated using inertial measurement unit (IMU) measurements with the aid of area states. The area states are in turn updated based on the position estimates. To avoid accumulated errors of IMU measurements, our algorithm uses WiFi received signal strength indicator (RSSI) to indicate the vicinity of the user to the routers. The experiment results show that our system can achieve satisfactory localization accuracy in a typical indoor environment.
\end{abstract}

\section{Introduction}
GPS-based localization systems can provide satisfactory performance in outdoor scenarios, while indoor localization systems still confront some challenges in terms of robustness, accuracy and responsiveness.
The reason for these challenges lies in the difficulty to obtain reliable measurements at low cost and to integrate multiple information sources effectively.
Many efforts have been made to study the indoor localization problem from different perspectives \cite{Radar,Tri,RSSIinaccuracy,Pedestrian,Foot,StepDetection,Network_localization_1,Network_localization_2}. Most of them can be categorized into two main categories, i.e., infrastructure-based and infrastructure-free approaches. Infrastructure-based approaches require to deploy wireless access points in the indoor environment and are performed on the basis of wireless measurements such as WiFi, UWB, and ultrasound. Among all these approaches, {WiFi-based ones attract much attention }since WiFi routers are ubiquitous and no additional infrastructure is needed for localization.  The WiFi-based approaches can be subdivided into two types: fingerprinting \cite{Radar} and ranging-based methods \cite{Tri}, both using RSSI to estimate users' positions. However, range-based methods often cause large localization errors due to the instability of WiFi signals in indoor environments and fingerprinting requires laborious off-line training and is not robust to environment changes \cite{RSSIinaccuracy}. Infrastructure-free approaches rely primarily on IMUs to locate a user by continuously estimating the displacement of positions from a known location, referred to as the dead-reckoning approach. A typical example is the pedestrian navigation system (PNS) \cite{Foot,Pedestrian,StepDetection}, which performs localization by detecting steps and estimating headings.
While most of these existing PNSs depend on high-quality IMU sensors \cite{Foot,StepDetection}, achieving satisfactory performance using the IMUs on smartphones is a challenging task.
%the IMUs on widely-available smartphones or have been implemented on smartphone platforms.
In addition to the measurements from WiFi routers and IMUs, an indoor map is also a valuable source of information for localization. Since walking patterns are partially determined by the geographical structure of the environments, the indoor map serves as the prior knowledge of space constraints in localization. A particle filter-based algorithm that combines map information and WiFi RSSI was proposed in \cite{MapPF}. In \cite{FootPath}, the algorithm improves the localization accuracy by matching the walking trajectories onto the map. To the best of the authors' knowledge, few studies have fully exploited map information in multi-sensor indoor localization.
%focused on the incorporation of the two mainstream information sources (WiFi and IMU) along with indoor map information. Moreover, there lacks a formal representation and full exploitation of indoor maps for indoor localization.
In this paper, we develop a localization algorithm fusing the information from IMUs, indoor maps as well as WiFi RSSI and implement a real-time system on Android smartphones.
We introduce the concept of \emph{area state} in the indoor map representation and develop different localization approaches according to the \emph{area states}. %provide an effective way to integrate \emph{area state} into the dead-reckoning approach. %Specifically,  are exploited  according to the area states. We also utilize the WiFi signals as vicinity indicators to refine the localization performance.
System implementation on Android platforms demonstrates the performance improvement achieved by exploiting map information.

\section{System Design}
\label{sec:sysdesign}
\begin{figure}[t]
\centering \includegraphics[width=0.9\linewidth]{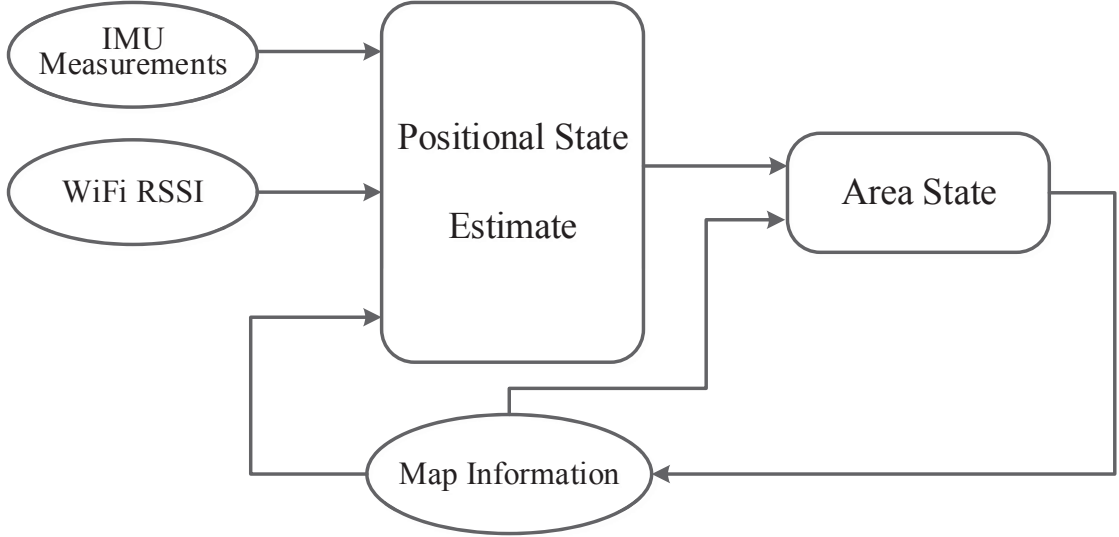}
\caption{Framework of the system.}\label{framework} \end{figure}\
The framework of our system is illustrated in Fig.~\ref{framework}.
In our system, the states of the user are categorized into two types: the positional state and the area state. The positional state refers to the coordinates of the user's position on a 2D map and the heading direction of the current step.
The area state refers to a particular area the user locates in, e.g., a certain corridor. At each step, the positional states are estimated based on IMU measurements with the aid of area states, and the area states are then updated by new positional states based on indoor map information. We also conduct experiments about WiFi RSSI in indoor environments and utilize the WiFi routers as \emph{vicinity indicators}.
More details about the algorithm are described as below.
\subsection{Dead-Reckoning using IMU Measurements} Although the accelerometer and gyroscope measurements on the phone are accurate according to our experiments, severe drift can hardly be avoided if the trajectories are obtained through double integration of the acceleration measurements. Therefore, we resort to the dead-reckoning approach based on detecting steps and step headings where the initial location and the step length of the user are assumed to be available.
Several regular patterns emerge in the accelerometer measurements when the user is walking with the smartphone in hand. We use the step detection method as in \cite{StepDetection} to detect the sharp drop of acceleration when the user is balancing out his/her steps.
%A step is counted when the acceleration magnitude falls more than the threshold within a sliding window period of time.
%magnetometer and gyroscope are generally used for real time localization. In indoor scenarios, however, magnetometer measurements are subject to severe interferences from electrical appliances and magnetic materials.
%Hence,
We estimate heading through integration of filtered angular velocity, which is not affected by the severe interferences from indoor electrical appliances and magnetic materials.
The initial heading and posture of the phone is assumed to be available at the beginning of the localization.
The turning can be detected by the abrupt change of the azimuth angle if the phone remains fixed relative to the user.
\subsection{Indoor Map Representation}
To the best of the authors' knowledge, there are few open-source geographical data of indoor maps. The incompleteness of geographical data makes it challenging to generalize map-based open-source projects to various sites.
Another challenge exists in the lack of standard representation rules to define all the elements in an indoor map.
Hence, we establish some intuitive and facile representation rules for our prototype, and verify the validity of our algorithm under these rules. \begin{table}[t]\footnotesize
\renewcommand\arraystretch{1.5}
\caption{Basic Elements in Formulated Indoor Maps} \centering \begin{tabular}{|c|c|c|c|} \hlinewd{1pt}
\textbf{Type} & \textbf{Area State} & \textbf{Element} & \textbf{Description} \\
\hline
\multirow{4}{*}{Node} & \multirow{2}{*}{Intersection} & Turning Point & Conjunction between corridors \\
\cline{3-4}
& & Door & Entrance of an open area \\
\cline{2-4}
& Open Area & Room \& Lobby & Relatively empty space \\
\cline{2-4}
& & Indicator & WiFi Router \\
\hlinewd{1pt} \multirow{4}{*}{Way} & Corridor & Corridor & Long and narrow passage \\
\cline{2-4}
& & Wall & Constraints enclosing a room \\
\cline{3-4}
& & DoorCorridor & Connection between door \& room \\
\cline{3-4}
& & WallCorridor & Connection between wall \& room \\
\hline
\end{tabular}
\label{Elements}
\end{table}

\label{sec:map_representation}
\begin{figure}[t]
\centering
\includegraphics[width=0.6\linewidth]{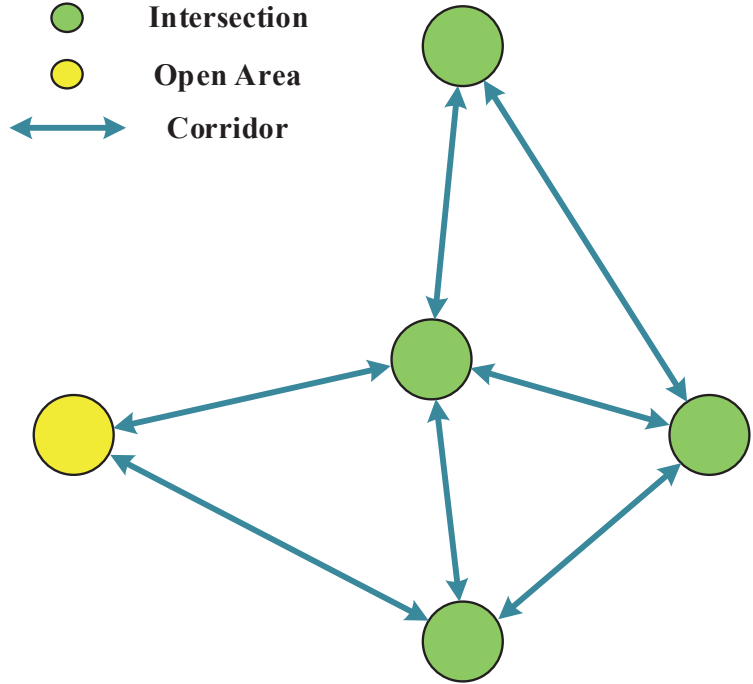}
\caption{The schematic diagram of area state representation.}\label{map_representation}
\end{figure}\
We establish our indoor maps from OpenStreetMap (OSM)\cite{OSM}, an open source geographical database.
Map data from OSM can be accessed as an eXtensible Markup Language (XML) file consisting of nodes, ways and relations, which can be annotated and labeled with key-value-pairs named as tags.
These tags allow us to set the rules by labeling the nodes as intersections and open areas, and the ways as corridors and walls.
All the basic elements in an indoor map are shown in Table \ref{Elements}.
According to the map structure, we divide the indoor environment into three main types of \emph{area states}: open areas, intersections and corridors.
The schematic diagram of area state representation on the map is presented in Fig. \ref{map_representation}.
\label{sec:state_transition_model}
\begin{figure}[t]
\centering
\includegraphics[width=0.8\linewidth]{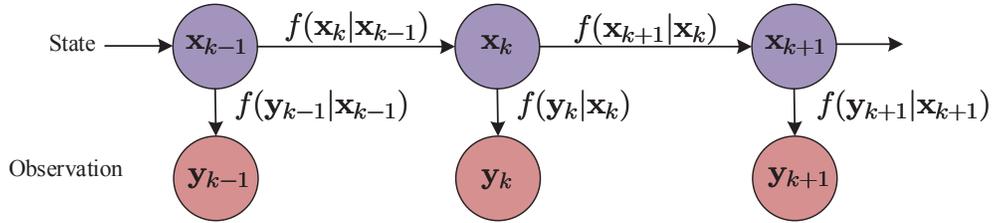}
\caption{Bayesian network representation of the variables in our algorithm.}\label{state_transition_model}
\end{figure}\

We model the time evolution of the joint positional and area states as well as the measurements as a hidden Markov model (HMM) shown in Fig. \ref{state_transition_model}. The positional state at step $k$ is denoted as $\left(p_{x,k},p_{y,k},\mathbf{h}_k^\text{T}\right)^\text{T}$, which contains the 2D coordinates on the map and the heading at step $k$. Specifically, the heading  of the user is represented by a unit direction vector $\mathbf{h}_k = {\left( \cos\theta_k,\sin\theta_k \right)}^\text{T}$, where $\theta_k$ denotes the heading angle.
We denote the area state of the user at step $k$ as $\alpha_k$.
Specifically, $\alpha_k$ determines the way of positional state transition from step $k$ to $k+1$, e.g., it is a 1D movement in the corridor or a 2D movement in an open area.
The joint state $\mathbf{x}_k=\left(p_{x,k},p_{y,k},\mathbf{h}_k^\text{T},\alpha_k\right)^\text{T}$ represents the combination of positional and area state at step $k$.
Meanwhile, the transition of area state results from the transition of positional state.
Accordingly, the observation of positional state at step $k$ is denoted as $\mathbf{y}_k=(p^o_{x,k},p^o_{y,k},{\mathbf{h}^o_k}^\text{T})^\text{T}$, where $\mathbf{h}^o_k$ is obtained by integrating the angular velocity measurements and $p^o_{x,k}$ and $p^o_{y,k}$ are obtained by moving a step length $s$ in the direction of $\mathbf{h}^o_k$, i.e., $p^o_{x,k} = s \cos \theta_k$, $ p^o_{y,k} = s \sin \theta_k$.
Based on $\mathbf{y}_k$, the position estimate is further refined by the map information if the user is aware of what area he/she is in. This is actually how we incorporate map information to mitigate noises and uncertainties of position estimate based on dead-reckoning. We elaborate the map-aided localization process as follows.

%and the joint probability density function of the observations and real positional states from step $1$ to step $K$ can be characterized as
%\begin{equation*}
%f(y_1,\cdots,y_K,x_1,\cdots,x_K)=\prod_{k=1}^{K}f(x_k|x_{k-1})f(y_k|x_k)
%\end{equation*}
%\textcolor[rgb]{1.00,0.00,0.00}{where $f(x_1|x_{0})=f(x_1)$.}

%As shown in Fig. \ref{state_transition_model}, the area state $\theta_k$ determines the way of positional state transition, e.g., it is a one dimension movement in the corridor or a two dimension movement in an open area.
%Furthermore, the observations are able to be filtered by the map information if the user is aware of what area he/she is in.
%This is actually how we incorporate map information to suppress noises and uncertainties of position estimate based on dead-reckoning.
%We elaborate the map-aided localization process as below.
\subsubsection{Walking in Corridor}
Since the width of a corridor is normally several meters and much shorter than its length, we assume that the user moves in one dimension when walking in the corridor. We set a threshold $\eta_\text{str}$ for the heading angle to determine whether the user is still walking in a corridor.
If so, the area state $\alpha_k$ remains to be \emph{Corridor} and the estimated heading is calibrated by the orientation information of the corridor.
\subsubsection{Turning}
If greater changes of heading angle than $\eta_\text{str}$ are detected and the estimated position is close to any turning point on the map, it is likely that the user is at a turning point.
The location information of the turning point can be used to calibrate the position estimate of the user.
Moreover, we determine which corridor the user is turning to by comparing the calculated heading and the bearing of the other $m$ corridors $c_{1},c_{2},\cdots,c_{m}$, connected with the turning point, whose bearings are $\mathbf{b}_{1},\mathbf{b}_{2},\cdots,\mathbf{b}_{m}$ accordingly, i.e. \begin{equation*}
\hat{c} = \argmax\limits_{c_{i},i=1,2,\cdots,m}  {\mathbf{h}^{o}_{k}}^\text{T} \mathbf{b}_{i}
\end{equation*}
%\begin{equation*}
%\hat{c} = \argmin\limits_{c_{i},i=1,2,\cdots,m}orienCom(h^{o}_{c},b_{i}) %\end{equation*}
%where $h^{o}_{c}$ is the current heading observation and the function $orienCom()$ returns the difference between two orientations.
The area state $\alpha_k$ is then updated as \emph{Intersection}. \subsubsection{Verifying the Turning}
If the next corridor detected is another long passage on the map, we introduce a Maximum a Posteriori (MAP) detection algorithm to verify the turning.
We do not align the current heading with the bearing of the next corridor instantly because chances are that the corridor selected is not the right track.

We model $\mathbf{h}^{o}_{j}$, the heading observation of the $j$-th step detected on a corridor, as a Gaussian distribution with the mean of $\mathbf{b}_{i}$ when the next corridor $c_{i}$ is given.
The variance of the Gaussian is set according to a number of experiments on walking patterns.
We use $l_j(c_i)=L({c_{i}|\mathbf{h}^o_{j-k+1},\mathbf{h}^o_{j-k+2},\cdots,\mathbf{h}^o_{j}})$ to represent the likelihood function of $c_{i}$ given observations from $k$ steps prior to $j$, $\mathbf{h}^o_{j-k+1},\mathbf{h}^o_{j-k+2},\cdots,\mathbf{h}^o_{j}$.
Furthermore, we assume $\mathbf{h}^o_{j-k+1},\mathbf{h}^o_{j-k+2},\cdots,\mathbf{h}^o_{j}$ are conditionally independent on $c_{i}$.
Then we obtain the likelihood function as
\begin{align*}
l_j(c_i)&=L({c_{i}|\mathbf{h}^o_{j-k+1},\cdots,\mathbf{h}^o_{j}})\\
&\propto p(\mathbf{h}^o_{j-k+1},\mathbf{h}^o_{j-k+2},\cdots,\mathbf{h}^o_{j}|c_{i})\\
&=p(\mathbf{h}^o_{j-k+1}|c_{i})\cdots p(\mathbf{h}^o_{j}|c_{i}) \end{align*}
As the likelihood evolves from step $j-k+1$ to step $j$, likelihood values of different hypotheses in $l_j$ will probably diverge, from which we can cut off the hypotheses with lower likelihood values and keep those with higher ones.
Thus, we are able to determine the next corridor $\hat{c}'$ by evaluating the likelihood function of consecutive $k$ step headings $\mathbf{h}^o_{j-k+1},\mathbf{h}^o_{j-k+2},\cdots,\mathbf{h}^o_{j}$ happen and make the hypothesis with the maximal likelihood the best guess, where we assume the prior probability of each corridor is equal, i.e., \begin{equation*}
\hat{c}' = \argmax\limits_{c_{i},i=1,2,\cdots,m}l_j(c_i)
\end{equation*} If $\hat{c}'$ is the same as the corridor estimated $\hat{c}$ in the first check, we align the current heading with the corridor's bearing.
Otherwise, we change the current corridor to $\hat{c}'$ and make $k$ steps' length compensation in distance.
Moreover, we will make another verification after $k$ steps until a right track is confirmed.
Then the area state $\alpha_k$ is updated as \emph{Corridor}.
\subsubsection{Walking in Open Area}
If the next corridor is a \emph{Door Corridor} that connects a door and a room or a lobby, the user is considered walking in the open area and so is the area state updated.
The user walks in a 2D area in this case with the space constraints imposed by the walls, i.e.,
the trajectories are not allowed to traverse the walls and the user can only walk along them if his/her trajectories cross the walls.
The entire process is summarized in Algorithm 1.
\begin{algorithm}[h]\label{algorithm1} \caption{Map Information Aided Algorithm} \begin{algorithmic}[1]
\begin{small}
\STATE \normalsize\textbf{Initialization:}\small \label{Alg:Initialization}
\STATE Given the initial phone posture, positional and area states, proper thresholds and step length.
\\\
\STATE \normalsize\textbf{Walk in corridor:}\small \label{Alg:Walk in a corridor}
\STATE Walk in 1D along the corridor.
\STATE \textit{\textbf{First Check:}} \label{Alg:First Check}
\STATE Detect the turning and calibrate the location with the turning points.
\IF {the user turns to a open space at a door} \STATE \textbf{go to:} \ref{Alg:Walk in open space}
\ENDIF
\STATE \textit{\textbf{Verify:}} \label{Alg:Double Check}
\STATE Perform MAP detection for the next corridor given consecutive $k$ headings.
\IF {the right track is confirmed}
\STATE Calibrate the heading with the corridor orientation.
\STATE \textbf{go to:} \ref{Alg:Walk in a corridor}
\ELSE
\STATE Change the current corridor and the last $k$ positional states.
\STATE \textbf{go to:} \ref{Alg:Double Check}
\ENDIF
%\IF {a turning is detected} \label{Alg:First Check}
%\IF {turning happens at a door}
%\STATE \textbf{go to:}\ref{Alg:Walk in open space}
%\ENDIF
%\ELSE
%\STATE \textbf{go to}\ref{Alg:Double Check}
%\ENDIF
%\STATE Double Check: {Perform Maximum A Posterior detection given consecutive $k$ headings.} \label{Alg:Double Check}
%\IF {the right track is confirmed}
%\STATE Calibrate the heading.\textbf{go to}\ref{Alg:Walk in a corridor}
%\ELSE
%\STATE \textbf{go to}\ref{Alg:Double Check}
%\ENDIF
\\\
\STATE \normalsize\textbf{Walk in open space:}\small \label{Alg:Walk in open space}
\STATE Walk freely in 2D area with walls and obstructions as space constraints.
\IF {the user turns to a corridor at a door}
\STATE \textbf{go to:} \ref{Alg:Walk in a corridor}
\ENDIF
\end{small}
\end{algorithmic}
\end{algorithm}

\subsection{WiFi Module}
In our system, WiFi module is used to refine the estimates of positional and area states.
We first conduct experimental studies on the propagation of WiFi signals and based on the experiment results, we show that it is suitable to use WiFi routers as vicinity indicators in our localization system.
\subsubsection{Variation of RSSI with distance}
\begin{figure}[t]
\centering
\includegraphics[width=0.8\linewidth]{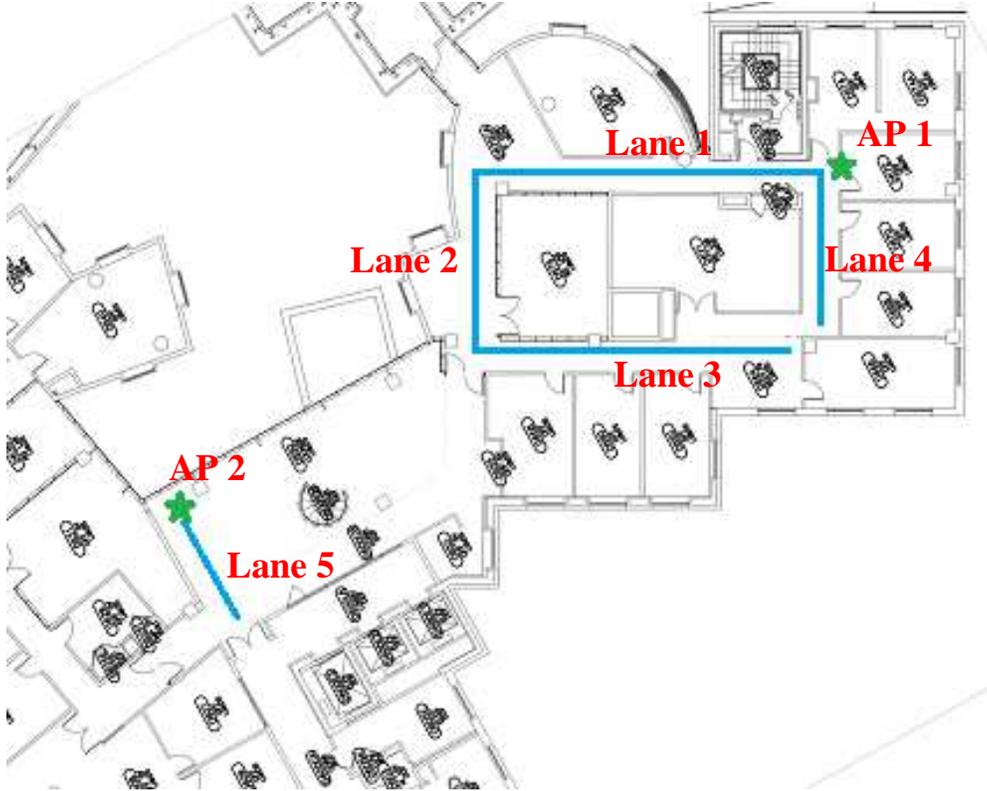}
\caption{Floor plan of 6-th floor of Dreyfoos Tower, Stata Center, MIT.}\label{fig:6thfloor}
\end{figure}

We conducted two experiments to investigate how RSSI varies with distance, when RSSI is measured on one Android smartphone. The experiments are conducted on 6th floor of Dreyfoos Tower, Stata Center, MIT. The floor plan are shown in Fig.~\ref{fig:6thfloor}. The WiFi routers we use work on 802.11ac standard. Meanwhile, we also consider the influence of tester's orientation {on RSSI}. In the first experiment, we measure the RSSI at a 0.3 m interval within a range of 4.5 m when the tester faces towards the WiFi router. In the second experiment, the tester is back against the router and all the other conditions are kept the same. The maximum distance is set to be 4.5 m, since we will use WiFi routers as indicators and this only requires the knowledge of WiFi propagation property within a short range. One hundred RSSI measurements are collected at each place in both experiments and they are obtained consecutively in about 30 s.

\begin{figure}[t]
\centering
\psfrag{Distance(m)}[][][.7]{Distance(m)}	
\psfrag{Router 1 Characteristic Plot}[][][0.6]{}
\psfrag{-55}[][][0.6]{-55}
\psfrag{-50}[][][0.6]{-50}
\psfrag{-45}[][][0.6]{-45}
\psfrag{-40}[][][0.6]{-40}
\psfrag{-35}[][][0.6]{-35}
\psfrag{-30}[][][0.6]{-30}
\psfrag{-25}[][][0.6]{-25}
\psfrag{0}[][][0.6]{0}
\psfrag{0.5}[][][0.6]{0.5}
\psfrag{1}[][][0.6]{1}
\psfrag{1.5}[][][0.6]{1.5}
\psfrag{2}[][][0.6]{2}
\psfrag{2.5}[][][0.6]{2.5}
\psfrag{3}[][][0.6]{3}
\psfrag{3.5}[][][0.6]{3.5}
\psfrag{4}[][][0.6]{4}
\psfrag{4.5}[][][0.6]{4.5}
\psfrag{Wifi RSSI(dBm)}[][][0.6]{WiFi RSSI (dBm)}
\psfrag{Samples when Facing the Router}[][][0.6]{\hspace{-4mm}Samples: facing the router}
\psfrag{Average when Facing the Router}[][][0.6]{\hspace{-4mm}Average: facing the router}
\psfrag{Samples when Backing to the Router}[][][0.6]{Samples: back towards the router}
\psfrag{Average when Backing to the Router}[][][0.6]{\hspace{-0.4mm}Average: back towards the router}
\includegraphics[width=1.0\linewidth]{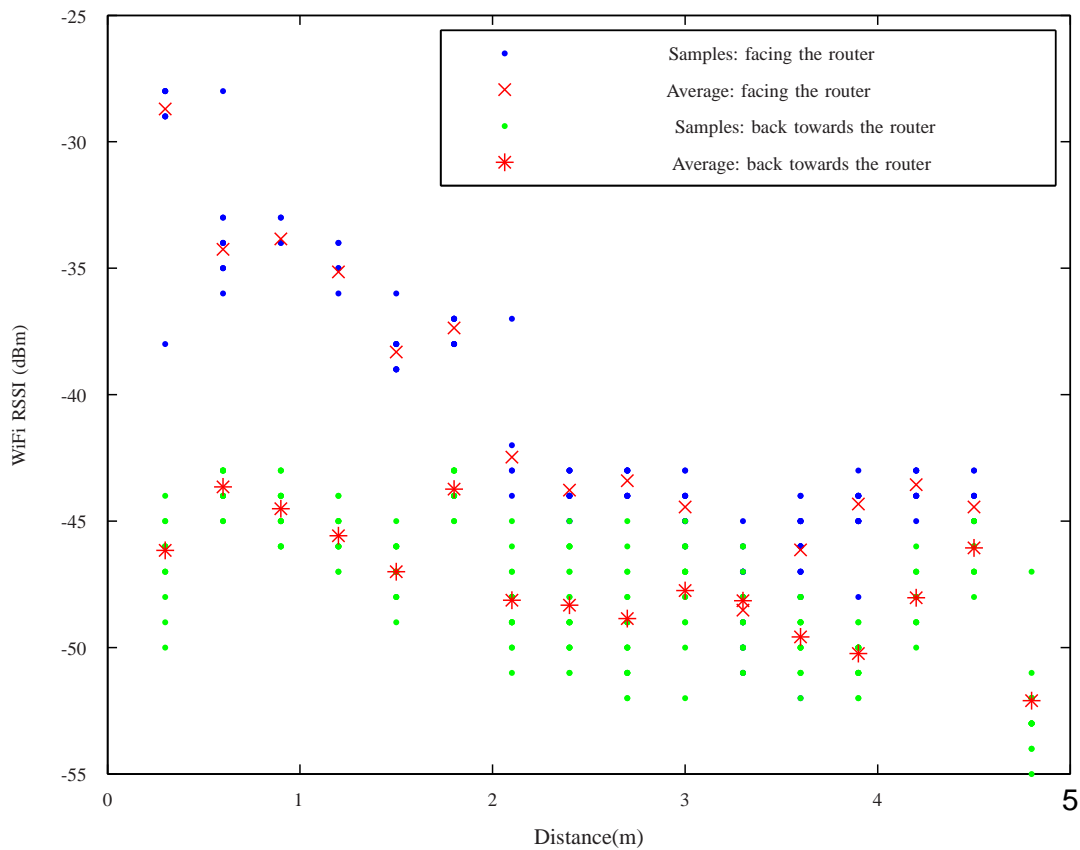}
\caption{Variation of RSSI as a function of the distance for AP 2.}\label{fig:RSSIdistance}
\end{figure}

Fig.~\ref{fig:RSSIdistance} depicts the samples and average values of WiFi RSSI measured at 15 different distances. When the tester faces towards the router, the RSSI is relatively sensitive to the distance change within a certain {distance} $\rho$ : a 35\% decrease from 0.3 m to 1.5 m,  while insensitive if the distance increases beyond $\rho \approx 2\text{m}$. In addition, RSSI sample values within and out of $\rho$ can be separated by a certain threshold $\phi$ ($-40$ dBm typically). This threshold is time-invariant and can be the same for routers of the same kind. {However, if the tester is back against the router and then the human body blocks the WiFi signal propagation, the RSSIs will be insensitive to distances in this case.} The experiment results are consistent with the well-known two-slope model \cite{two_slope}.

\
\
\subsubsection{Human body effects on WiFi signals}
We also investigate human body effects on WiFi signals. The RSSI is measured and collected under four different scenarios: (1) mobile phone left alone on the cart, (2) mobile phone placed on the cart and the user stands by, (3) mobile phone held in hand and the user walks towards the router, (4) mobile phone held in hand and the tester walks away from the router. The distance between mobile phone and the router is around 18 m.

\begin{figure}[t]
\centering
\psfrag{RSSI/dBm}[][][.7]{RSSI/dBm}
\psfrag{Frequency}[][][.7]{Frequency}
\psfrag{0}[][][0.6]{0}
\psfrag{0.2}[][][0.6]{0.2}
\psfrag{0.4}[][][0.6]{0.4}
\psfrag{0.6}[][][0.6]{0.6}
\psfrag{0.8}[][][0.6]{0.8}
\psfrag{1}[][][0.6]{1}
\psfrag{-50}[][][0.6]{-50}
\psfrag{-60}[][][0.6]{-60}
\psfrag{-70}[][][0.6]{-70}
\psfrag{-80}[][][0.6]{-80}
\psfrag{Scenario 1}[][][.7]{Scenario 1}
\psfrag{Scenario 2}[][][.7]{Scenario 2}
\psfrag{Scenario 3}[][][.7]{Scenario 3}
\psfrag{Scenario 4}[][][.7]{Scenario 4}
\includegraphics[width=1\linewidth]{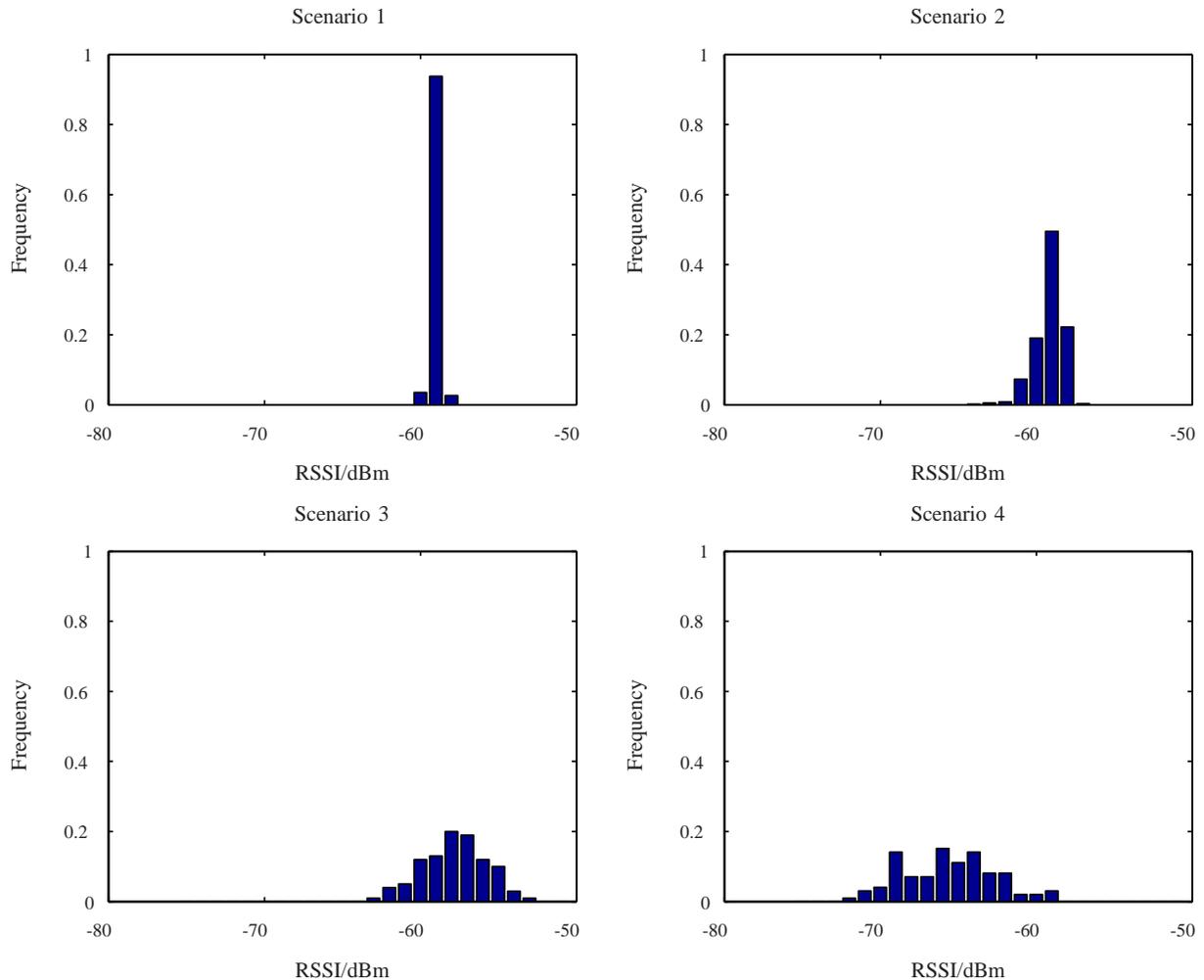}
\caption{Statistical distribution of the RSSI values under 4 scenarios.}
\label{fig:humaninterference}
\end{figure}

The statistics of RSSI under each scenario are presented in Fig. \ref{fig:humaninterference} and Table \ref{table:huminterference}. First, by comparing scenario 1 and scenario 2-4, we can find that the variance of RSSI is mainly caused by human body and other interferences, e.g. interferences from other devices, are much less significant.  Second, by comparing scenario 3 and 4, we see that human body can cause great RSSI loss if obstructing the direct path of signal propagation. We conclude that human body effects make the WiFi signals unstable. In fact, unstable WiFi signals may lead to large errors on WiFi-based localization, especially when the user and the WiFi router are not close enough. Therefore, it is hard to obtain accurate position estimates if we use the WiFi signals to directly estimate the distance between the user and the router.

\begin{table}[t]
\renewcommand\arraystretch{1.5}
\caption{Mean and variance of RSSI in 4 different scenarios.}
\centering
\begin{tabular}{|c|c|c|c|c|}
\hline
$\text{Scenario}$ &1 & 2 & 3 & 4 \\
\hline
$\text{Mean of RSSI/dBm}$ &-59.05 & -59.22  & -57.83 & -65.71  \\
\hline
$\text{Variance of RSSI/dBm}^{2}$ &0.46 & 1.95  & 4.22 & 9.45 \\
\hline
\end{tabular}

\label{table:huminterference}
\end{table}

\subsubsection{WiFi router as a vicinity indicator}
The experiment results in Fig. \ref{fig:RSSIdistance} show that RSSI is above a certain threshold only if the user is within a short range of a WiFi router. This implies that a high RSSI indicates vicinity to a certain router. Therefore, our system uses the RSSI in the way that if multiple RSSI sample values from certain router are above $\phi$, {we consider that }the phone is within a distance of $\rho$ from that
router and the positional states is calibrated using the router's location. This can avoid the deterioration of localization accuracy caused by unstable WiFi signals and also reduce the accumulative errors of IMU measurements.

\section{Implementation and Evaluation}
\label{sec:eval}
In this section, we elaborate more details about the implementation and evaluate the performance of the on-line localization system.
\subsection{Experiments}
All the experiments are conducted on the smartphone SamSung Galaxy S4, GT-I9500 running the Android operating system 4.4.3.
%The reference coordinate system of the phone is shown in Fig. \ref{coordinates}.
Our algorithm is performed and evaluated on the 6-th floor of Stata Center Dreyfoos Tower at MIT.
\subsubsection{IMU Measurements}

\begin{figure}[t]
	\centering
	% \hspace{-5mm}
	\subfigure[]{
    \psfrag{time(s)}[][][.7]{time}	
    \psfrag{Step detection plot}[][][0.6]{}
    \psfrag{0.85}[][][0.6]{0.85}
    \psfrag{0.9}[][][0.6]{0.9}
    \psfrag{0.95}[][][0.6]{0.95}
    \psfrag{1}[][][0.6]{1}
    \psfrag{1.05}[][][0.6]{1.05}
    \psfrag{1.1}[][][0.6]{1.1}
    \psfrag{10}[][][0.6]{10}
    \psfrag{10.5}[][][0.6]{10.5}
    \psfrag{11}[][][0.6]{11}
    \psfrag{11.5}[][][0.6]{11.5}
    \psfrag{12}[][][0.6]{12}
    \psfrag{12.5}[][][0.6]{12.5}
    \psfrag{13}[][][0.6]{13}
    \psfrag{13.5}[][][0.6]{13.5}
    \psfrag{14}[][][0.6]{14}
    \psfrag{14.5}[][][0.6]{14.5}
    \psfrag{acceleration magnitude(g)}[][][0.6]{acceleration magnitude(g)}
    \psfrag{acceleration magnitude}[][][0.6]{\hspace{-4mm}acceleration magnitude}
    \psfrag{steps detected}[][][0.6]{\hspace{-4mm}steps detected}
    \includegraphics[width=0.8\columnwidth,draft=false]{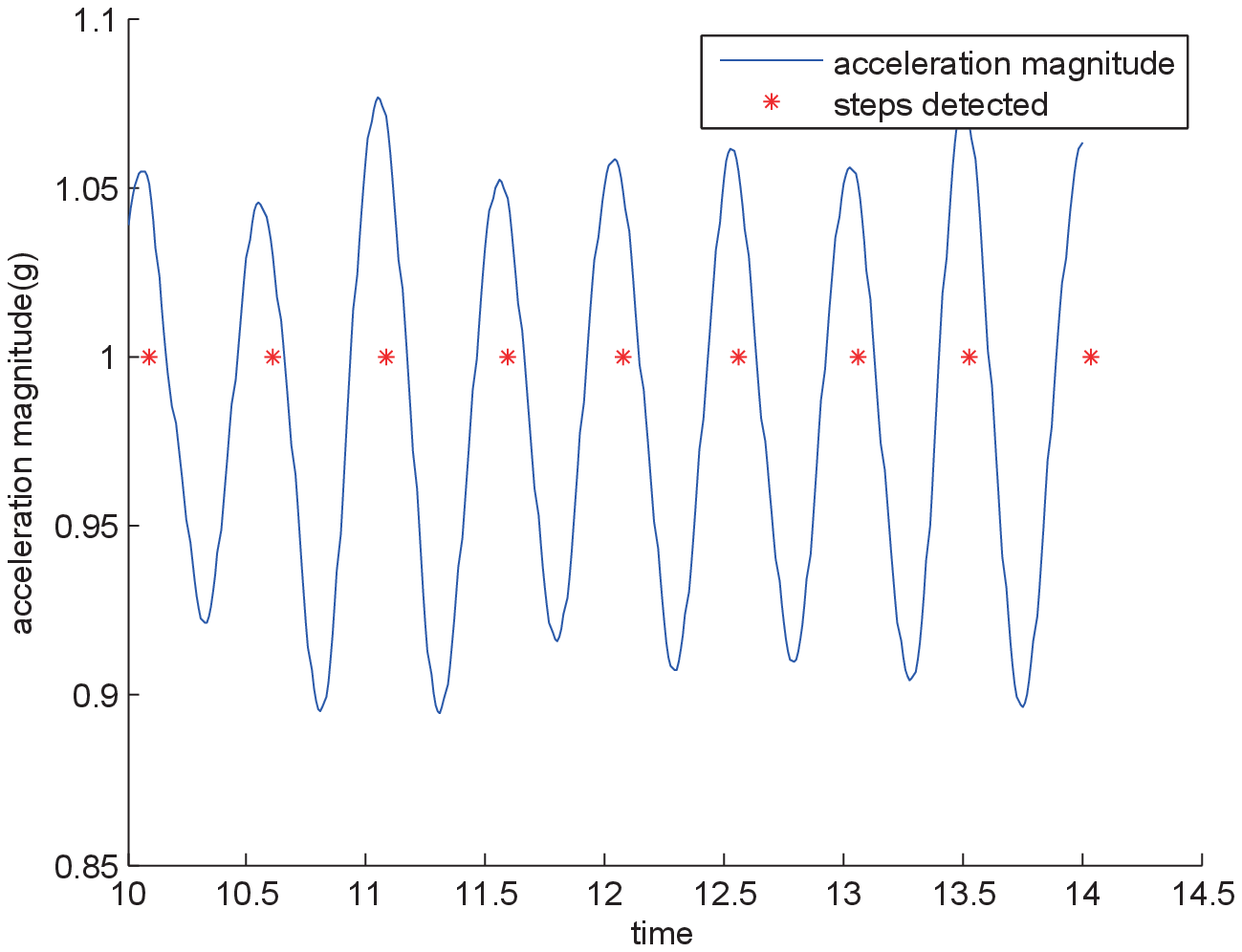}
     }
	\hspace{-3mm}
	\subfigure[]
	{
	\psfrag{time(s)}[][][.7]{time(s)}	
    \psfrag{Heading estimation plot}[][][0.6]{}
    \psfrag{-400}[][][0.6]{-400}
    \psfrag{-350}[][][0.6]{-350}
    \psfrag{-300}[][][0.6]{-300}
    \psfrag{-250}[][][0.6]{-250}
    \psfrag{-200}[][][0.6]{-200}
    \psfrag{-150}[][][0.6]{-150}
    \psfrag{-100}[][][0.6]{-100}
    \psfrag{-50}[][][0.6]{-50}
    \psfrag{0}[][][0.6]{0}
    \psfrag{50}[][][0.6]{50}
    \psfrag{10}[][][0.6]{10}
    \psfrag{20}[][][0.6]{20}
    \psfrag{30}[][][0.6]{30}
    \psfrag{40}[][][0.6]{40}
    \psfrag{60}[][][0.6]{60}
    \psfrag{70}[][][0.6]{70}
    \psfrag{80}[][][0.6]{80}
    \psfrag{degree}[][][0.6]{degree}
    \psfrag{Azimuth estimation after bias is removed}[][][0.6]{\hspace{-4mm}Azimuth estimation before bias removed}
    \psfrag{Azimuth estimation after bias is removed}[][][0.6]{\hspace{-4mm}Azimuth estimation after bias removed} \includegraphics[width=0.8\columnwidth,draft=false]{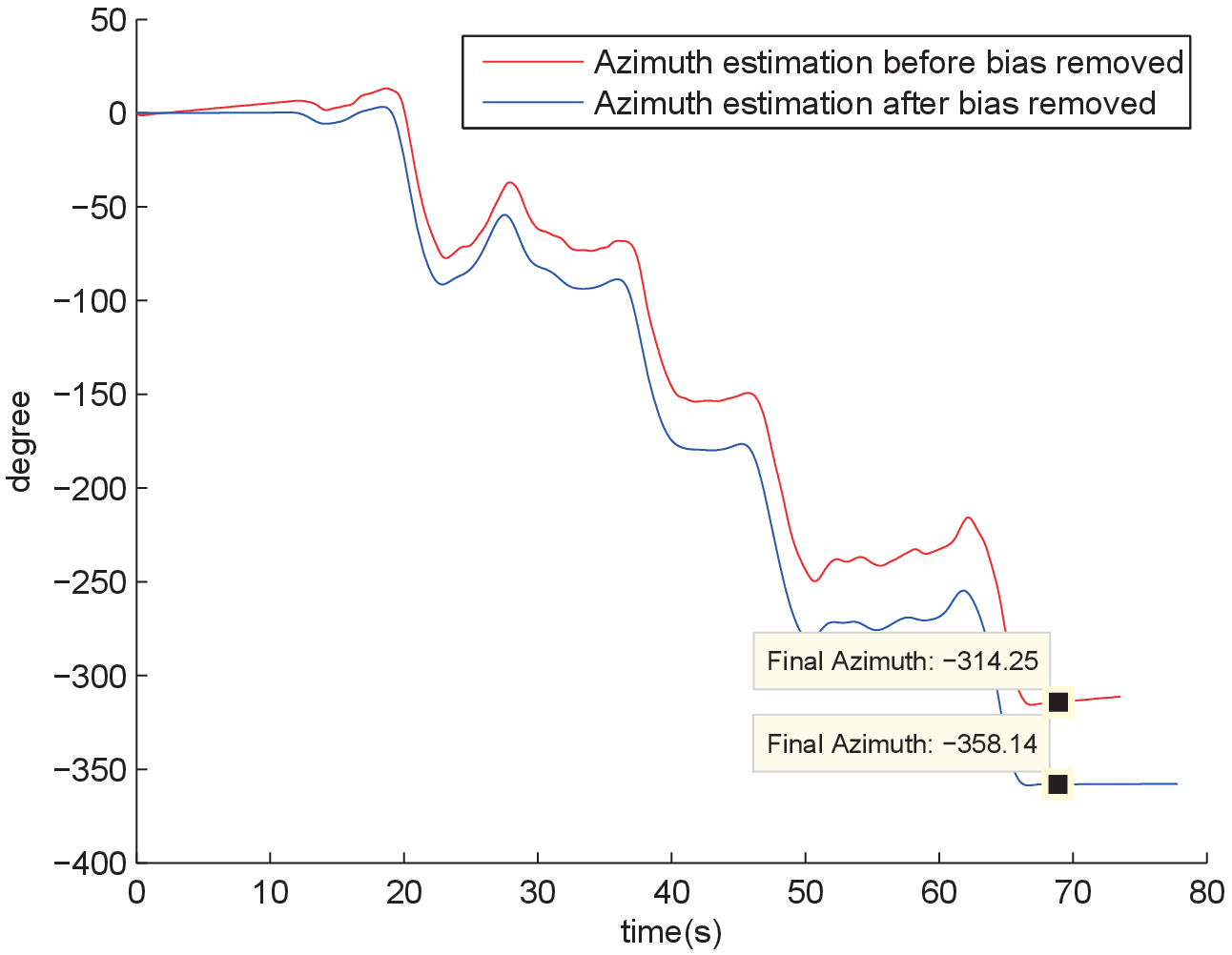}
	}
	\caption{Step detection (a) and heading estimation (b).}
	\label{fig:heading_stepdetection}
\end{figure}

%\begin{figure}[ht]
%%\begin{tabular}{cc}
%\psfrag{10}[][][0.6]{0}
%\centering
%\psfrag{10}[][][0.6]{0}
%\begin{minipage}{0.8\linewidth}\psfrag{10}[][][0.6]{0}
%  \centerline{\psfrag{10}[][][0.6]{0}
%  \psfrag{10}[][][0.6]{0}
%\includegraphics[width=\textwidth]{heading_stepdetection1.eps}
%\psfrag{10}[][][0.6]{0}}
%  \centerline{\scriptsize{(a) Step Detection}
%  \psfrag{10}[][][0.6]{0}
%  \psfrag{10}[][][0.6]{0}}
%  \psfrag{10}[][][0.6]{0}
%\end{minipage}
%\hfill
%\begin{minipage}{0.8\linewidth}
%  \centerline{\includegraphics[width=\textwidth]{heading_stepdetection2.eps}}
%  \centerline{\scriptsize{(b) Heading Estimation}}
%\end{minipage}
%\caption{Step detection and heading estimation results after preprocessing.}
%\label{fig:heading_stepdetection}
%\end{figure}\
For IMU experiments, we assume the user is holding the phone in hand with the Y-axis roughly aligning with the heading and Z-axis pointing up to the sky while walking.
Small jolting or shaking can be well-compensated by our area state-aided algorithm.
%We test the quality of the IMU measurements on the smartphone SamSung Galaxy S4 in terms of noise level, resolution and sampling rate.
We also exploit Butterworth low pass filter to smooth the raw data and subtract the average of the first measurements to remove the bias.

In our experiments, the smartphone is first held in hand. The variations of the acceleration magnitude during walking are exploited to detect the steps. Fig. \ref{fig:heading_stepdetection} (a) implies the effectiveness of step detection.
Then, the smartphone is fixed flat on a cart. We push the cart along the \emph{Lane 4, 3, 2}, and \emph{1}.
It is shown in Fig. \ref{fig:heading_stepdetection} (b) that the variations of azimuth estimates after a circle are way below $360$ degrees due to the bias.
The estimation errors can be reduced to less than $2$ degrees after bias removal.
\subsubsection{Map Representation Process}
We employ JOSM, an open-source map editing software, and the floor plan, to generate the map data in XML format.
The floor map is required to be calibrated both in orientation and scale at first.
Then some significant feature points, such as turning points and doors are labeled% so that the corridors and trajectories are identified.
and the relations of the map elements pairs, e.g., doors and rooms, are represented by the links between them.
The absolute positions and orientations of the elements can be further calculated from the longitude and latitude of each node.
In general, the process of aligning and labeling the map is very easy to operate and the wiki-style editing makes it possible to construct map collaboratively.
%\begin{figure}[t]
%	\centering
%	% \hspace{-5mm}
%	\subfigure[]{
%\psfrag{Reference Point}[][][0.6]{Reference Point}
%\psfrag{Anchor}[][][0.6]{~~~~~~~~~~~~~~WiFi Router}
%    \includegraphics[width=0.8\columnwidth,draft=false]{RP_AP.eps}
%      \centerline{\scriptsize{(a) Anchors and Reference Points}}
%     }
%	\subfigure[]
%	{
%\psfrag{Ground truth}[][][.6]{\hspace{1mm}~~Ground truth}
%\psfrag{With area state/anchor}[][][0.6]{~With area state \& indicator}
%\psfrag{With anchor}[][][0.6]{~~~~~With indicator}
%\psfrag{With area state}[][][0.6]{With area state}
%\psfrag{Only IMU}[][][0.6]{~~~~~~Only IMU}
% \includegraphics[width=0.8\columnwidth,draft=false]{trace2.eps}
%   \centerline{\scriptsize{(b) Experiment Trajectories}}
%	}
%	\caption{Anchors, reference points and experiment trajectories.}
%	\label{fig:groundtruth1}
%\end{figure}
\begin{figure}[ht]
%\begin{tabular}{cc}
\centering
\begin{minipage}{0.8\linewidth}
\psfrag{Reference Point}[][][0.6]{Reference Point}
\psfrag{Anchor}[][][0.6]{~~~~~~~~~~~~~~WiFi Router}
  \centerline{\includegraphics[width=0.9\textwidth]{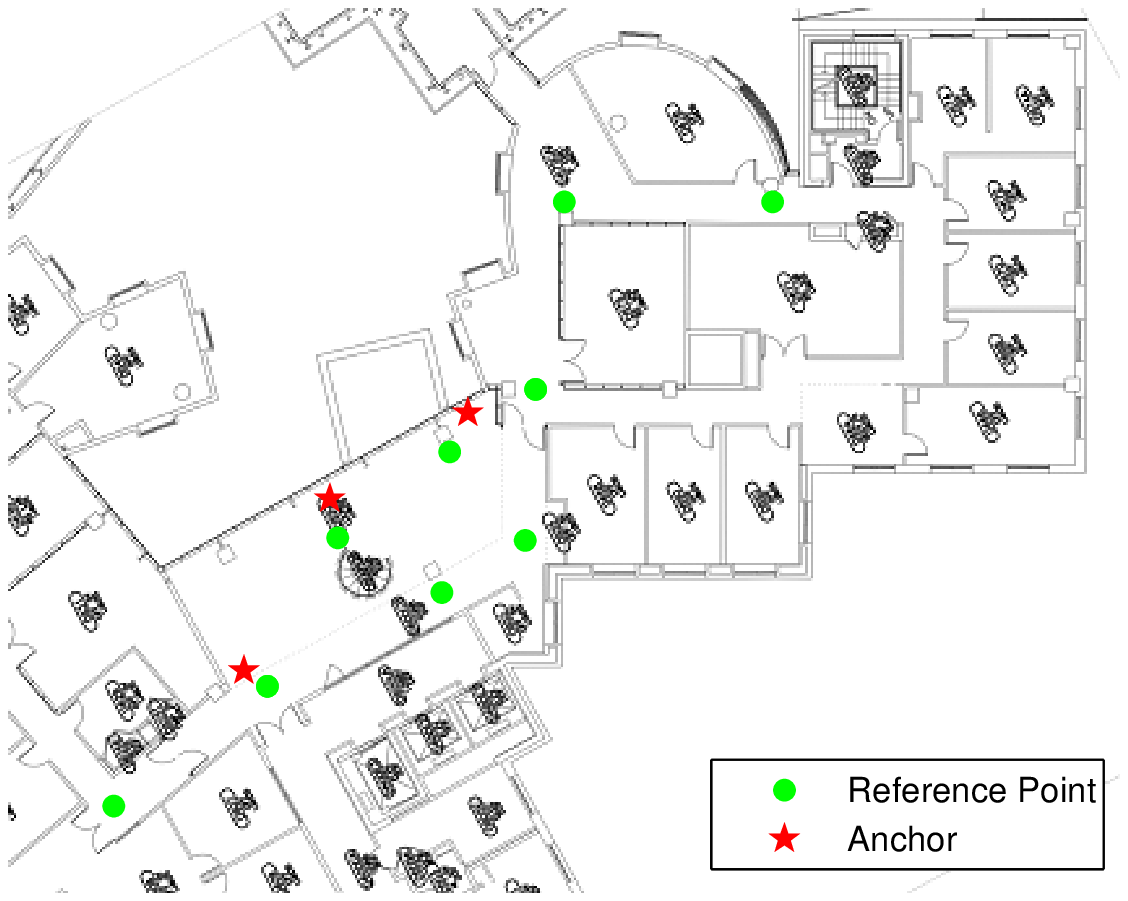}}
  \centerline{\scriptsize{(a) Anchors and Reference Points}}
\end{minipage}

\begin{minipage}{0.8\linewidth}
\psfrag{Ground truth}[][][.6]{\hspace{1mm}~~Ground truth}
\psfrag{With area state/anchor}[][][0.6]{~With area state \& indicator}
\psfrag{With anchor}[][][0.6]{~~~~~With indicator}
\psfrag{With area state}[][][0.6]{With area state}
\psfrag{Only IMU}[][][0.6]{~~~~~~Only IMU}
  \centerline{\includegraphics[width=0.9\textwidth]{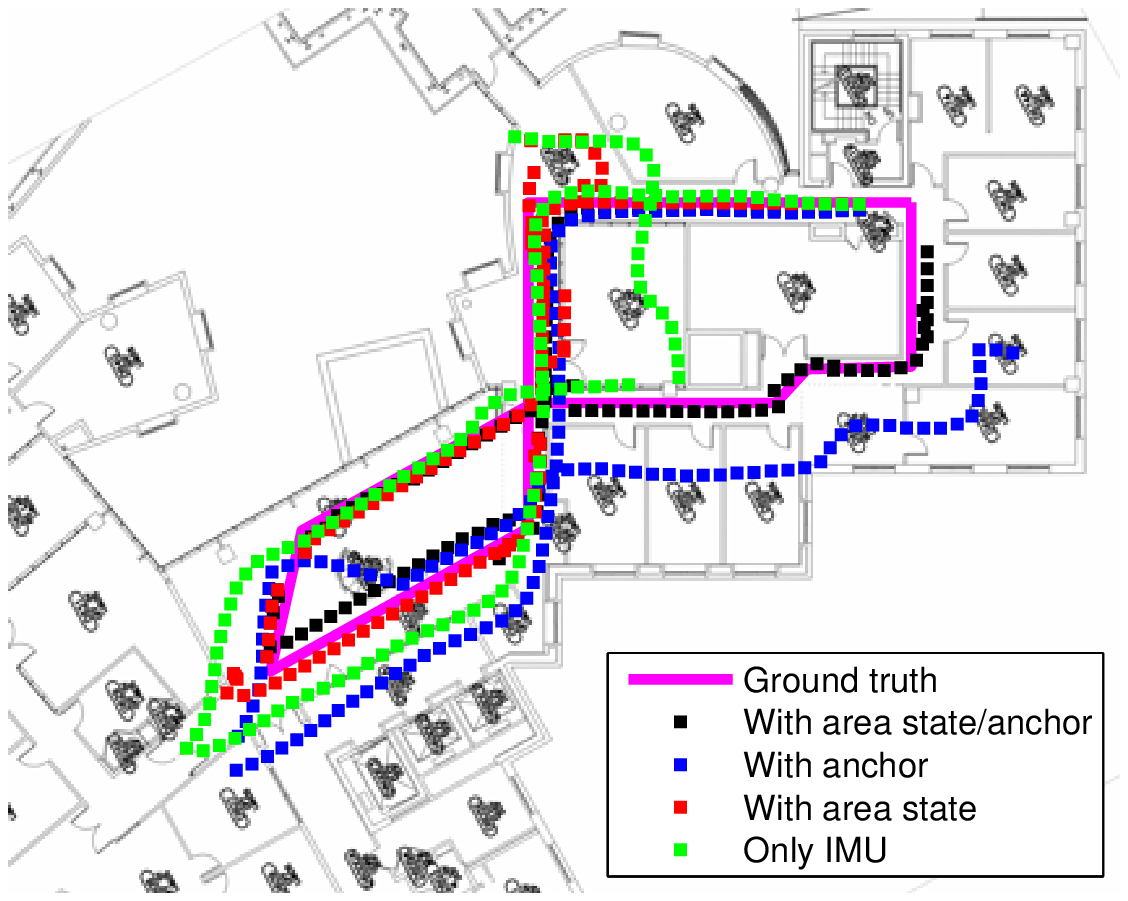}}
%  \centerline{\scriptsize{(b) Experiment Trajectories}}
\end{minipage}
\caption{Anchors, reference points and experiment trajectories.}
\label{fig:groundtruth1}
\end{figure}\
\begin{figure}[t]
\centering
\psfrag{0}[][][0.6]{0}
\psfrag{0.1}[][][0.6]{0.1}
\psfrag{0.2}[][][0.6]{0.2}
\psfrag{0.3}[][][0.6]{0.3}
\psfrag{0.4}[][][0.6]{0.4}
\psfrag{0.5}[][][0.6]{0.5}
\psfrag{0.6}[][][0.6]{0.6}
\psfrag{0.7}[][][0.6]{0.7}
\psfrag{0.8}[][][0.6]{0.8}
\psfrag{0.9}[][][0.6]{0.9}
\psfrag{1}[][][0.6]{1}
\psfrag{2}[][][0.6]{2}
\psfrag{4}[][][0.6]{4}
\psfrag{6}[][][0.6]{6}
\psfrag{8}[][][0.6]{8}
\psfrag{10}[][][0.6]{10}
\psfrag{12}[][][0.6]{12}
\psfrag{CDF}[][][0.6]{CDF}
\psfrag{Localization Error (m)}[][][0.6]{Localization Error (m)}
\psfrag{With area state/anchor}[][][0.6]{~~With area state \& indicator}
\psfrag{With anchor}[][][0.6]{~~~With indicator}
\psfrag{With area state}[][][0.6]{With area state}
\psfrag{Only IMU}[][][0.6]{~~~~Only IMU}
\includegraphics[width=0.8\linewidth]{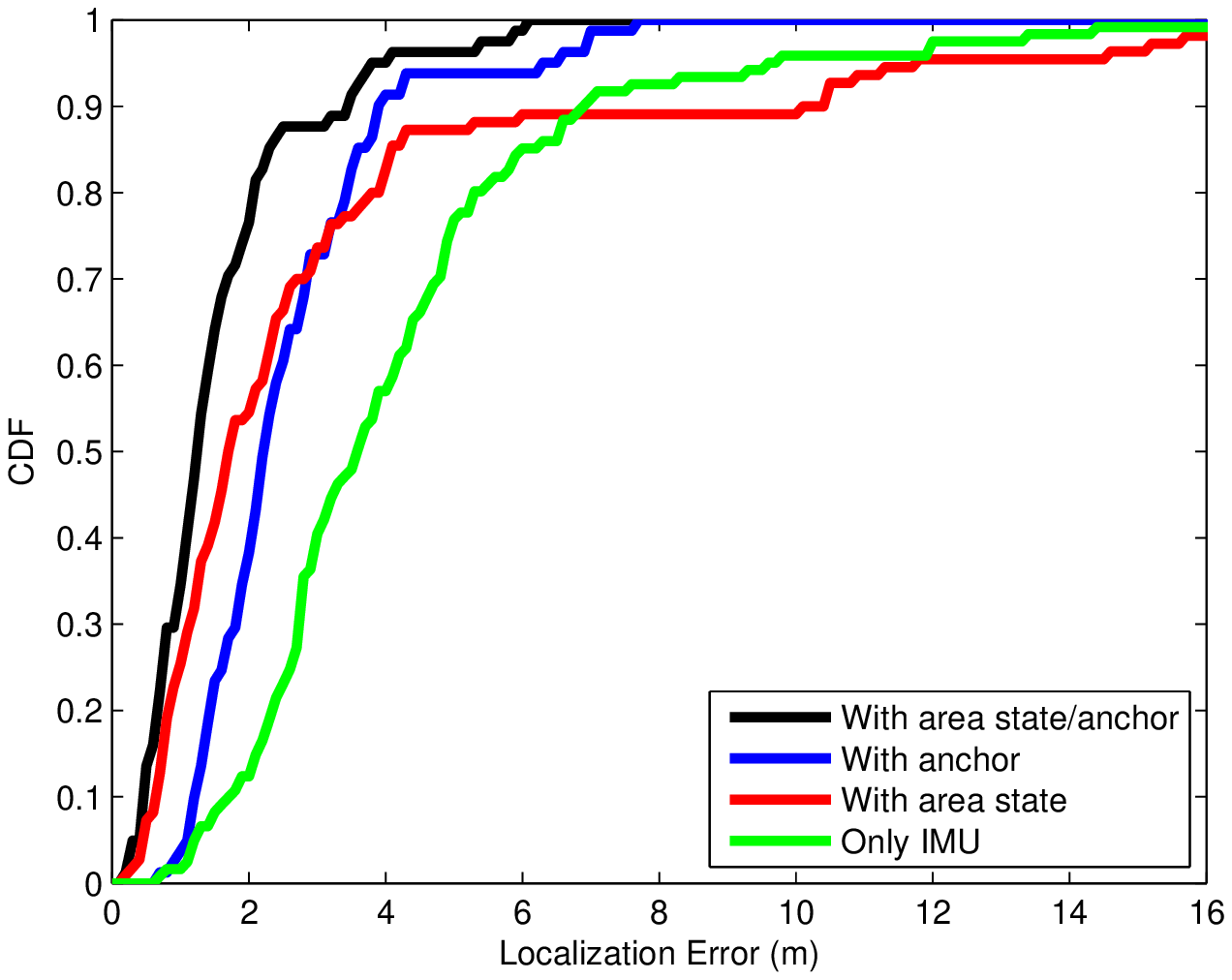}
\caption{Cumulative distribution function of localization errors.}\label{CDF}
\end{figure}\
\subsection{Performance Analysis}
To evaluate the performance of our localization system, a user walks around the 6th floor of Stata Center with the phone in hand.
Three WiFi routers are deployed around the lobby and their locations are marked on the map as shown in Fig. \ref{fig:6thfloor}.

As illustrated in Fig. \ref{fig:groundtruth1}, the real-time estimation of trajectories are compared with the ground truth to justify the validity of our algorithm.
It shows that our system presents robust performance in corridors, which attributes primarily to map constraints.
The position estimates are calibrated at turning points despite of some step misdetections or step length estimation errors.
Moreover, constant corrections of headings with the assistance of corridors can well-compensate for the small attitude variations of the phone, which allows \emph{painless} gestures of users during localization.
In open area, the trajectories are constrained by the walls and obstructions and refined by the WiFi indicators.
WiFi routers deployed at doors can also be exploited to detect the position transitions of different map elements, e.g., from the corridor to the lobby or vice verse.
The localization errors of our algorithm at the reference points, shown in Fig.~\ref{fig:6thfloor}, are recorded after ten repetitions of experiments. Fig. \ref{CDF} plots the cumulative distribution function (CDF) of localization errors using different algorithms. It can be seen that the proposed algorithm outperforms those algorithms not using either the area states or the WiFi indicators. In particular, the average error is reduced from 4.2 meters to 1.6 meters. Moreover, map information and WiFi indicators provide different types of performance gains in terms of localization errors. Map information can effectively mitigate small errors due to measurements calibration while cause large errors when no absolute positional information, e.g., WiFi indicators, is exploited. Nonetheless, the position refinement from WiFi indicators is susceptible to small localization errors if the threshold $\phi$ is not properly selected.

% localization accuracy of meter-level is achieved in our algorithm, which outperforms the results without the aid of area states.
\section{Conclusions}
We designed and implemented a real-time indoor localization algorithm by leveraging measurements from IMUs and WiFi RSSI with map information. The proposed algorithm can effectively exploit map information and achieve desirable localization accuracy with the aid of area states. Our localization algorithm also uses WiFi RSSI to indicate the vicinity of the user to the router. This system can serve as a prototype for the design of advanced indoor localization system in the future.

\end{document}